\documentclass[aps,prd,twocolumn,showpacs,nofootinbib]{revtex4-1}

\usepackage{amssymb} \usepackage{amsmath} \usepackage{graphicx}
\usepackage{epsfig,latexsym}
\usepackage{mathtools}% http://ctan.org/pkg/mathtools
\RequirePackage{xspace} \allowdisplaybreaks

\begin{document}

\def\bef{\begin{figure}}
\def\eef{\end{figure}}

\newcommand{\nl}{\nonumber\\}

\newcommand{\ans}{ansatz }
\newcommand{\be}[1]{\begin{equation}\label{#1}}
\newcommand{\beq}{\begin{equation}}
\newcommand{\ee}{\end{equation}}
\newcommand{\beqn}[1]{\begin{eqnarray}\label{#1}}
\newcommand{\eeqn}{\end{eqnarray}}
\newcommand{\bd}{\begin{displaymath}}
\newcommand{\ed}{\end{displaymath}}
\newcommand{\mat}[4]{\left(\begin{array}{cc}{#1}&{#2}\\{#3}&{#4}
\end{array}\right)}
\newcommand{\matr}[9]{\left(\begin{array}{ccc}{#1}&{#2}&{#3}\\
{#4}&{#5}&{#6}\\{#7}&{#8}&{#9}\end{array}\right)}
\newcommand{\matrr}[6]{\left(\begin{array}{cc}{#1}&{#2}\\
{#3}&{#4}\\{#5}&{#6}\end{array}\right)}
\newcommand{\cvb}[3]{#1^{#2}_{#3}}
\def\lsim{\raise0.3ex\hbox{$\;<$\kern-0.75em\raise-1.1ex
e\hbox{$\sim\;$}}}
\def\gsim{\raise0.3ex\hbox{$\;>$\kern-0.75em\raise-1.1ex
\hbox{$\sim\;$}}}
\def\abs#1{\left| #1\right|}
\def\simlt{\mathrel{\lower2.5pt\vbox{\lineskip=0pt\baselineskip=0pt
           \hbox{$<$}\hbox{$\sim$}}}}
\def\simgt{\mathrel{\lower2.5pt\vbox{\lineskip=0pt\baselineskip=0pt
           \hbox{$>$}\hbox{$\sim$}}}}
\def\unity{{\hbox{1\kern-.8mm l}}}
\newcommand{\eps}{\varepsilon}
\def\ep{\epsilon}
\def\ga{\gamma}
\def\Ga{\Gamma}
\def\om{\omega}
\def\omp{{\omega^\prime}}
\def\Om{\Omega}
\def\la{\lambda}
\def\La{\Lambda}
\def\al{\alpha}
\newcommand{\ov}{\overline}
\renewcommand{\to}{\rightarrow}
\renewcommand{\vec}[1]{\mathbf{#1}}
\newcommand{\vect}[1]{\mbox{\boldmath$#1$}}
\def\tm{{\widetilde{m}}}
\def\mcirc{{\stackrel{o}{m}}}
\newcommand{\Dm}{\Delta m}
\newcommand{\dm}{\varepsilon}
\newcommand{\tanb}{\tan\beta}
\newcommand{\nbar}{\tilde{n}}
\newcommand\PM[1]{\begin{pmatrix}#1\end{pmatrix}}
\newcommand{\up}{\uparrow}
\newcommand{\down}{\downarrow}
\def\omE{\omega_{\rm Ter}}

%
%%%%%%%%%%     mauri    %%%%%%%%%%%%%%%%%%%%%%%%%%%%%%%%%

\newcommand{\Dsusy}{{susy \hspace{-9.4pt} \slash}\;}
\newcommand{\DCP}{{CP \hspace{-7.4pt} \slash}\;}
\newcommand{\mc}{\mathcal}
\newcommand{\gr}{\mathbf}
\renewcommand{\to}{\rightarrow}
\newcommand{\gtc}{\mathfrak}
\newcommand{\wh}{\widehat}
\newcommand{\br}{\langle}
\newcommand{\kt}{\rangle}

%%%%%%%%%%%%%%%%%%%%%%%%%%%%%%%%%%%%%%%%%%%%%%%%%%%%%%%%%%

% barbara Ricci  %definizione di minore e maggiore simile
\def\lsim{\mathrel{\mathop  {\hbox{\lower0.5ex\hbox{$\sim$}
\kern-0.8em\lower-0.7ex\hbox{$<$}}}}}
\def\gsim{\mathrel{\mathop  {\hbox{\lower0.5ex\hbox{$\sim$}
\kern-0.8em\lower-0.7ex\hbox{$>$}}}}}
%%%%%%%%%%%%%%%%%%%%%%%%%%%%%%%%%%

\def\nn{\\  \nonumber}
\def\de{\partial}
\def\brf{{\mathbf f}}
\def\bbf{\bar{\bf f}}
\def\bF{{\bf F}}
\def\bbF{\bar{\bf F}}
\def\bA{{\mathbf A}}
\def\bB{{\mathbf B}}
\def\bG{{\mathbf G}}
\def\bI{{\mathbf I}}
\def\bM{{\mathbf M}}
\def\bY{{\mathbf Y}}
\def\bX{{\mathbf X}}
\def\bS{{\mathbf S}}
\def\bb{{\mathbf b}}
\def\bh{{\mathbf h}}
\def\bg{{\mathbf g}}
\def\bla{{\mathbf \la}}
\def\bmu{\mathbf m }
\def\by{{\mathbf y}}
\def\bmu{\mbox{\boldmath $\mu$} }
\def\bsig{\mbox{\boldmath $\sigma$} }
\def\bunity{{\mathbf 1}}
\def\cA{{\cal A}}
\def\cB{{\cal B}}
\def\cC{{\cal C}}
\def\cD{{\cal D}}
\def\cF{{\cal F}}
\def\cG{{\cal G}}
\def\cH{{\cal H}}
\def\cI{{\cal I}}
\def\cL{{\cal L}}
\def\cN{{\cal N}}
\def\cM{{\cal M}}
\def\cO{{\cal O}}
\def\cR{{\cal R}}
\def\cS{{\cal S}}
\def\cT{{\cal T}}
\def\eV{{\rm eV}}

%
%%%%%%%%%%%%%%%%%%%%%%%%%%%%%%%%%%%%%

\title{Way-out to the Gravitino problem in intersecting D-brane Pati-Salam models}

\author{Andrea Addazi$^1$}\email{andrea.addazi@infn.lngs.it}
\affiliation{$^1$ Dipartimento di Fisica,
 Universit\`a di L'Aquila, 67010 Coppito AQ and
LNGS, Laboratori Nazionali del Gran Sasso, 67010 Assergi AQ, Italy}

\author{Maxim Yu Khlopov$^2$}\email{khlopov@apc.univ-paris7.fr}
\affiliation{$^2$ Centre for Cosmoparticle Physics Cosmion;
National Research Nuclear University MEPHI (Moscow Engineering Physics 
Institute), Kashirskoe Sh., 31, Moscow 115409, Russia;
and APC laboratory
10, rue Alice Domon et Leonie Duquet 75205 Paris Cedex 13, France}

\begin{abstract}

We discuss the gravitino problem 
in contest of the Exotic see-saw mechanism for neutrinos and Leptogenesis,
UV completed by intersecting D-branes Pati-Salam models. 
In the Exotic see-saw model, supersymmetry is broken at high scales 
$M_{SUSY}>10^{9}\, \rm GeV$ and this seems in contradiction 
with gravitino bounds from inflation and baryogenesis. 
However, if gravitino is the Lightest Stable Supersymmetric Particle, 
it will not decay into other SUSY particles, avoiding the gravitino problem and providing a good Cold  Dark Matter.
Gravitini are Super Heavy Dark Particles and they can be produced by non-adiabatic expansion 
during inflation. Intriguingly, from bounds on the correct abundance of dark matter, 
we also constrain the neutrino sector. We set a limit on the exotic instantonic coupling 
of $<10^{-2}\div 10^{-3}$. This also sets important constrains on the Calabi-Yau compactifications
and on the string scale.  This model strongly motivates 
very high energy DM indirect detection of 
neutrini 
and photons of $10^{11}\div 10^{13}\, \rm GeV$: gravitini can decay on them in a cosmological time
because of soft R-parity breaking effective operators.

\end{abstract}

\maketitle
\section{Introduction}

Recently, we suggested a new coherent model for leptogenesis and neutrino mass 
in contest of IIA superstring theory \cite{Addazi:2015yna}, inspired 
by a model previously suggested in Ref.\cite{Addazi:2015hka}. 
In particular, we discussed a simple intersecting D-branes construction reproducing a Pati-Salam 
model in the low energy limit. This proposal is a reincarnation of the old see-saw type I mechanism, 
which remains the simplest idea in order to unify neutrino mass with leptogenesis
\footnote{Let us also mention that a recent $SO(10)$ leptogenesis analysis was proposed in Ref.\cite{DiBari:2015svd}.
On the other hand, leptogenesis scenari related to radiative neutrino masses and Dirac see-saw mechanisms were considered in 
 Refs.\cite{Lu:2016ucn,Gu:2016hxh}. }. 
But, contrary to $SO(10)$ inspired model, the main contributions to the RH neutrino
mass matrix come from non-perturbative stringy effects 
known in literature as exotic stringy instantons \footnote{In other contributions, 
we show that exotic instantons can generate an effective Majorana
mass for the neutron, with testable implications in neutron-antineutron physics
 \cite{Addazi:2015hka,Addazi:2014ila,Addazi:2015ata,Addazi:2015rwa,Addazi:2015fua,Addazi:2015ewa,Addazi:2015goa,Addazi:2016xuh}
. }.
In IIA superstring theory, these effects can be calculated and controlled, 
providing a precise prediction on the neutrino mass matrix
  \cite{Blumenhagen:2006xt,Ibanez1,Ibanez2}. 
In our model, we assume two main hypothesis:
i) the non-perturbative corrections to RH mass matrix from exotic instantons
dominate with respect to standard mass terms coming from Pati-Salam 
spontaneous symmetry breaking pattern;
ii) Supersymmetry and Left-Right symmetry are assumed to be spontaneously broken 
at high scale, i.e. $M_{SUSY}, M_{LR}\geq 10^{9}\, \rm GeV$.
Relaxing the (i)-hypotesis and assuming non-perturbative corrections to be sub-leading, 
we can just obtain a similar scenario to standard $SO(10)$ one.
However, this scenario is in tension with Davidson-Ibarra (DI) bound $M_{R}^{DI}\simeq 10^{9}\, \rm GeV$ \cite{DI}.
The DI mass bound is set by a correct leptogenesis mechanism, 
considering 
 first generation RH neutrino decaying into leptons and Higgs. 
The DI bound is 
 usually avoided assuming 
degeneracies among RH neutrino masses.
But such an assumption is a fine-tuning.
In particular, we remark that the popular $SO(10)$ PS really
is in tension with DI-bound:
the lightest RH neutrino is hierarchically constrained to be $M_{RH1}<<10^{9}\, \rm GeV$, 
i.e. it is not a good candidate for see-saw leptogenesis. As a consequence, 
in $SO(10)$ PS we necessary have to assume an accidental degeneracy 
among RH neutrini really not understood in their symmetry breaking pattern. 
On the other hand, in our scenario, exotic instantons can easily generate {\it democratically }
RH neutrino masses at high scale. So that, the coincidence 
of masses is dynamically recovered by exotic instantonic processes,
 easily avoiding DI bounds. 
The (ii)-hypothesis  is also indirectly  motivated by recent results of LHC: TeV-MSSM and TeV-Left-Right scenari 
seem to be rule-out by recent data. 
On the other hand, the (ii)-assumption eliminates a lot of undesired free-parameters in our model, 
rendering it constrainable by low energy observables and leptogenesis consistency. 
For instance, supersymmetric particles are not produced by RH neutrini decays. 
These assumptions reduced our model to $13$ free parameters
mainly parametrizing our ignorance on the internal Calabi-Yau compactification. 
All parameters are also reduced by the rigidity of Pati-Salam symmetry, 
relating the CKM matrix of quarks to the Pontecorvo matrix of neutrini. 
However, they can be constrained by $11$ low energy observables in 
neutrino and quark physics and the leptogenesis consistency. 
From these $11$ in-puts, we are able to predict a precise range for 
the $\theta_{13}\neq 0$ oscillation angle. 
The precise determination of $\theta_{13}$-angle is 
a hot topic of neutrino physics and it is crucially important 
 to constrain it in future, for our understanding of 
baryogenesis. 
As we will show, we will relate measures in neutrino and quark physics with the topology and geometric shape
parameters of the Calabi-Yau compactification.
However the (ii)-hypothesis can lead to the gravitino problem.
The gravitino problem set a bound of $m_{\tilde{G}}<10^{6}\, \rm TeV$
\cite{Khlopov:1984pf}. So that a generic supersymmetric scenario with $M_{SUSY}>10^{9}\, \rm GeV$
can lead to several problems. For instance, the gravitino could be unstable and 
rapidly decay into supersymmetric partners and Pati-Salam particles, 
essentially leading to a disastrous washing-out of our leptogenesis scenario. 
In this paper, we discuss the simplest way-out to this problem, 
saving our leptogenesis picture. 
It regards the gravitino stability: if the gravitino is a LSP particle 
in the SUSY spectrum, then we will be safe by gravitino reheating decays.
Of course, for $M_{SUSY}>10^{9}\,\rm GeV$, gravitino is expected to be a very heavy 
particle. However, it can have a mass small as $m_{\tilde{G}}\simeq 10^{-9}\times M_{SUSY}$
as for standard gravitino warm dark matter in MSSM.
But it is also possible a scenario $m_{\tilde{G}}\simeq (10^{-1}\div 1)\, M_{SUSY}$
with $m_{\tilde{G}}<m_{SUSY-particles}$. Of course, if stable, they 
can also be more massive than gravitino problem bound. 
This offers the intriguing possibility to solve and complete our leptogenesis mechanism 
relating it with a candidate for Cold Dark Matter (CDM). 
In particular, we will discuss a scenario in which 
gravitini are Super Heavy Dark Matter (SHDM) particles 
gravitationally produced by the non-adiabatical expansion during inflation epoch. 
We obtain new consistency bounds from CDM
abundance. In particular, we will put emphasis on the 
new bounds set to neutrino sector from CDM limits. 

This paper is organized as follows:
in subsections A and B we review the main aspects of Exotic see-saw from 
the theoretical and phenomenological side;
in Section II, we discuss our main results on SHDM-gravitino production 
during inflation; in Section III, we show our conclusions. 

\subsection{Exotic see-saw mechanism in (un)oriented Pati-Salam quivers }

Let us review the basic aspects of the exotic see-saw model proposed in Ref.\cite{Addazi:2015yna}. 
The complete set-up of our model is shown in Fig.1 of Ref. \cite{Addazi:2015yna}.
In the low energy limit,
our model is described by a PS gauge group $U(4)\times Sp(2)_{L}\times Sp(2)_{R}$:
  $U(4)$ is generated by a stacks of 4 D6-branes and their images
$U'(4)$, identified under a $\Omega$-plane;
  $Sp(2)_{L,R}$ are supported on two stacks of two D-branes each
lying on top of the $\Omega$-plane
  We also consider two Euclidean $D2$-branes (or $E2$-branes)
  on top of the $\Omega$-plane, corresponding to 
two Exotic $O(1)$ Instantons. Let us call these as $E2',E2''$.    
Quarks and leptons in Left and Right  fundamental representations $F_{L,R}\equiv {\bf 4}_{L,R}$, are reproduced as excitations 
 of open strings attached to the $U(4)$-stack and the Left or Right $Sp(2)_{L,R}$-stacks (respectively). 
 Analogously, 
 Higgses $\bar{H}\equiv {\bf 4}_{R}$ and its conjugate $H$ are introduced as 
 extra intersections of $U(4)$-stack with $Sp(2)_{R}$.
 Extra color states $\Delta=(10,1,1)$, and their conjugates, are obtained as excitations of open strings attached 
 to the $U(4)$-stack and its mirror image $U(4)'$-stack. 
 $\phi_{LL}=(1,3,1)$ and $\phi_{RR}=(3,1,1)$ correspond to
strings attached on the $Sp(2)_{L,R}$ with both end-points  (respectively).
 Higgs fields $h_{LR}=(2,2,1)$ are massless strings attached to $Sp(2)_{L}$ and $Sp(2)_{R}$. 
 The quiver on the left of Fig.1 automatically encodes the following super-potential terms
\cite{Addazi:2015hka}:
 \be{WYuk}
 \mathcal{W}_{Yuk}=Y^{(0)}h_{LR} F_{L}F_{R}+\frac{Y^{(1)}}{M_{1}}F_{L}\phi_{LL}F_{L}\Delta \ee
 $$+\frac{Y^{(2)}}{M_{2}}F_{R}\phi_{RR}F_{R}\Delta^{c}+\frac{Y^{(3)}}{M_{3}}h_{LR}\phi_{RR}h_{RL}\phi_{LL}+\mu h_{LR}h_{RL}$$
 $$+Y^{(5)}h_{LR}F_{L}\bar{H}+\frac{Y^{(6)}}{M_{6}}F_{R}\phi_{RR}\bar{H}\Delta^{c}$$
 $$ +
\frac{Y^{(7)}}{M_{7}}F_{L}F_{L}F_{R}F_{R}+\frac{Y^{(8)}}{M_{8}}F_{L}F_{L}\bar{H}\bar{H}+\frac{Y^{(9)}}{M_{9}}F_{L}F_{L}F_{R}\bar{H}$$
  \be{WH}
 \mathcal{W}_{H}=m_{\Delta}\Delta\Delta^{c}+\frac{1}{4 M_{4}}(\Delta\Delta^{c})^{2} \ee
 $$+\frac{1}{2}m_{L}\phi_{LL}^{2}+\frac{1}{2}m_{R}\phi_{RR}^{2}+\frac{1}{3!}a_{L}\phi_{LL}^{3}$$
 $$+\frac{1}{3!}a_{R}\phi_{RR}^{3}
+\mu'H\bar{H}+\frac{Y^{(10)}}{M_{10}}\bar{H}\phi_{RR}\bar{H}\Delta^{c}+\mathcal{Y}^{(11)}HH\Delta$$
$Y^{(...)}$ are Yukawa matrices;
while the mass scales $M_{...}$ are considered as free parameters. 
In fact mass scales depend 
 on the particular completion of our model:
they can be near the string scale $M_{S}$ as well as at lower scales
On the other hand, mass terms $m_{\Delta}$ and $m_{L,R}$ can be generated
by R-R or NS-NS 3-forms fluxes in the bulk,
in a T-dual Type IIB description: 
$m_{\Delta}\sim\Gamma^{ijk}\langle\tau H_{ijk}+iF_{ijk}\rangle$,
$m_{L,R}\sim\Gamma^{ijk}\langle\tau H^{(L,R)}_{ijk} + iF^{(L,R)}_{ijk}~\rangle$,
with $H_3$ RR-RR and $F_3$ NS-NS 3-forms.
The 
Super-potential terms (\ref{WE2p},\ref{WE2})
can be generated by two $E2$-brane instantons,
with Chan-Paton group $O(1)'$. 
They intersect twice the $U(4)$ stack 
and $O(1)''$ intersects one time $Sp(2)_{R}$-stack and once the $U(4)$-stack. 

We remind that ermionic modulini $\tau_{i},\tau'_{i},\omega'_{\alpha}$
 are massless excitations 
of open strings ending on $U(4)-O(1)$, $U(4)-O(1)'$,
$Sp(2)_{R}-O(1)'$ respectively, where
$i=1,4$ and $\alpha=1,2$ are indices of $U(4)$ and $Sp(2)_{R}$ respectively. 
So that, integrating 
Integrating over the fermionic modulini, we exactly 
recover the interactions (\ref{WE2p}).
In particular, the dynamical scales generated in 
(\ref{WE2})
are 
 $\mathcal{M}_{0}'=Y^{'(1)}M_{S}e^{+S_{E2'}}$
 and  $\mathcal{M}_{0}''=Y^{''(1)}M_{S}e^{+S_{E2''}}$,
 where 
$S_{E2',E2''}$ depend on geometric moduli, associated to 3-cycles of the $CY_{3}$, around which $E2',E2''$ 
are wrapped.

The spontaneous/St\"uckelberg breaking pattern down to the (MS)SM (minimal supersymmetric standard model) is
\be{pattern}
U(4)\times Sp(2)_{L} \times Sp(2)_{R}\rightarrow_{\langle Stu \rangle} SU(4) 
\times Sp(2)_{L} \times Sp(2)_{R}
\ee
$$ \rightarrow_{\langle \bar{H},H,h \rangle} SU(3) \times Sp(2)_{L}\times U(1)_{Y}$$
($St$ for St\"uckelberg).
 $h_{LR}$ contain the standard Higgses for the standard electroweak symmetry breaking.

The extra $U(1)_{4}\subset U(4)_{c}$ 
is anomalous in gauge theory. 
But  anomalous abelian gauge group can be cured 
by
a generalization of the Green-Schwarz mechanism. 
Usually, this mechanism necessary requires 
generalized Chern-Simons 
(GCS) terms and 
new massive St\"uckelberg vector boson $Z'$ associated to $U(1)_{4}$.

So that quiver nodes are split 
$4\rightarrow 3+1$ and $2_{R}\rightarrow 1+1'$
and we obtain a new effective Higgsed quiver shown in Fig.1-(b) of Ref. \cite{Addazi:2015yna}.  
In the new effective quiver, 
we consider the intersections of a new exotic instanton 
$E2$, intersecting once $U(1)$ and once $\hat{U}(1)'$
-
where $\hat{U}'(1)$ indicates the $\Omega$-image of $U'(1)$. 
So that, an extra non-perturbative mass matrix term is obtained: 
\be{WE2p}
\mathcal{W}_{E2'}= {1\over 2} \mathcal{M}_{ab}'N^{a}_{R}N^{b}_{R}
\ee
where $N_{R}^{a}$ are RH neutrini, $a=1,2,3$ label neutrino species. 
$N_{R}$ were contained inside $F_{R}$ as singlet components. 
In particular, the mass matrix has a structure 
$\mathcal{M}_{ab}=Y_{ab}^{(0)'}M_{S}e^{-S_{E2'}}$, where
$Y_{ab}^{(4)}$ is the Yukawa matrix.
The Yukawa matrix parametrizes 
 masses and mixings among RH neutrini. 
 The mass matrix also depends on 
on the particular $E2$ intersections 
with ordinary D6-branes stacks. 
However, this sets a hierarchical bound
on the Left-Right symmetry:
the superpotential (\ref{WE2p}) can be generated only after 
spontaneous symmetry breaking of Pati-Salam group:
 $U(4)_{c} \rightarrow U(3)_{c}$, and $Sp(2)_{R}\rightarrow U'(1)$.

The electroweak symmetry breaking in our model is obtained by $\langle h_{LR} \rangle$ of the complex Higgs bi-doublets 
$h_{LR}$. This leads to the important
 the tree-level mass relations
\be{treelevel}
m_{d}=m_{e}\,\,\,\,\,{\rm and}\,\,\,\,\,\,m_{u}=m_{D}
\ee
( $m_{D}$ are Dirac masses of neutrini).
From (\ref{treelevel}), tight hierarchy constraints on RH neutrini masses
are predicted: as a result the neutrino hierarchy is related to the
up-quark's one. It is interesting to observe that the hierarchy obtained at the perturbative level
(with closed-string fluxes generating the $M_{2}$ scale)
is corrected by exotic instantons, 
parametrized by $\mathcal{M}_{ab}$.  
Left-Right symmetry breaking pattern implies 
\be{wh}
V_{L}=V_{CKM}\,\,\,\,{\rm and}\,\,\,\,\,\,m_{D}=m_{u}
\ee
where $V_{CKM}$ is the Cabibbo-Kobayashi-Maskawa matrix.
The mass matrix has a structure
\be{MassMatrix}
M= \left( \begin{array}{cc} 0 & m_{D}
\ \\ m_{D} & M_{R} \ \\
\end{array} \right) 
\ee
RH neutrino masses have two contributions
$$M_{R}=M_{R}^{P}+M_{R}^{E2'}$$
where 
$$M_{R}^{P} =\langle\phi_{RR}\rangle\langle S^{c} \rangle/M_{2}$$
and 
$$M_{R}^{E2'}=\mathcal{M}_{ab}'$$ 
This implies that light neutrino mass matrix $m_{\nu}$
\be{ssf}
m_{\nu}\simeq -m_{D}\left(M_{R}^{P}+M_{R}^{E2}\right)^{-1}m_{D}
\ee
Then,
the inverted see-saw formula is
\be{issf}
M_{R}=M_{R}^{p}+M_{R}^{E2}\simeq -m_{D}m_{\nu}^{-1}m_{D}
\ee
using the matrix symmetry $m_{D}=m_{D}^{T}$.

Rel.(\ref{issf}) provides the information 
on the RH neutrino mass matrix $M_{R}$, 
extracted using: 
i)  LH neutrino mass matrix
$m_{\nu}$, ii) a PS-quark-lepton symmetry.

A natural situation for model is that 
$E2'$ induces non-perturbative mass terms for RH 
neutrini of the same order:
 $$M^{E2}_{R,1}\simeq M^{E2}_{R,2} \simeq M^{E2}_{R,3}$$ 
 where 1,2,3 are generation indeces. As a consequence, 
$M^{E2}_{R,1,2,3}\simeq 10^{9}\div 10^{10}\ \rm GeV$ and
we obtain a highly  degenerate RH mass spectrum
This does not imply a highly degenerate LH mass spectrum: a large quark-lepton hierarchy 
is contained in Dirac matrix $m_{D}$. 

Let us comment that the low energy limit a 
IIA SUGRA 
plus non-perturbative non-perturbatively generated superpotentials 
are obtained by our IIA superstring theory model. 
The scalar potential in perturbative supergravity 
has a supersymmetric form 
\be{form}
V_{F}={\rm}\left(\frac{K}{M_{Pl}^{2}} \right)\left\{(K^{-1})^{i\bar{j}}D_{i}WD_{\bar{j}}W^{*}-\frac{|W|^{2}}{M_{Pl}^{2}}\right\}
\ee
where the second term disappears in the global rigid susy limit  $M_{Pl}\rightarrow \infty$. 
The second term comes from the auxiliary fields 
in the gravity multiplet. 
The corresponding gravitino mass term is related to the K\"ahler and super-potentials as
\be{gravmass}
\mathcal{L}_{g.m.}={\rm exp}\left(\frac{K}{2M_{Pl}^{2}} \right)\frac{W^{*}}{M_{Pl}^{2}}\psi_{a}\sigma^{ab}\psi_{b}
\ee
In SUGRA the goldstino is 
eaten by the gravitino 
becoming massive. 
So that, in the limit of $M_{Pl}^{2}\rightarrow \infty$, 
the gravitino is massless, as also explicitly 
obtained performing such a limit in (\ref{gravmass}). 
In flat space-time vacua, is rigid SUSY is unbroken 
the gravitino mass must be zero, 
related to the condition $V_{F}=0 \rightarrow F=0$. 
On the other hand, in AdS vacua, 
the F-term can be null even allowing a non-zero scalar potential
and consequently a massive gravitino. 
In fact the supersymmetric algebra in AdS is different
by SuperPoincar\'e group algebra. 
Supersymmetry in AdS relates superpartners 
without mass degeneracy.   
However, Rels.(\ref{form})-(\ref{gravmass}) can be affected by non-perturbative stringy effects 
not present at all in standard supergravity. 
For instance,  the scalar potential and the perturbative vacua can be lifted by R-R and NS-NS supersymmetric 
fluxes in the Calabi-Yau compactification, consequently lifting the gravitino mass. 
An example is provided in Ref. \cite{Kachru:2003aw}, where in a T-dual IIB context 
supersymmetric fluxes are introduced in order to stabilize Calabi-Yau moduli. 
On the other hand, also the presence of non-supersymmetric R-R and NS-NS three-forms 
fluxes in the bulk can generate an extra soft-susy mass term for the gravitino,
as $m_{3/2}^{NP} \psi_{a} \sigma^{ab}\psi_{b}$, where $m_{3/2}^{NP}\sim \langle \tau H+iF \rangle$ (flux indices contractions omitted). 
In our scenario, 
the gravitino has to be the Lightest Supersymmetric Particles. 
So that
we assume that non-perturbative fluxes will not be higher than the 
supersymmetric scale and superpartners masses.

\subsection{Region of parameters and main results}

Under model generalities discussed above, 
In Ref. \cite{Addazi:2015yna} we consider the following assumptions: 

\vspace{0.4cm}
I) Supersymmetry and Left-Right symmetry are spontaneously broken at high scales:
$M_{SUSY}, M_{LR}>M_{R}\simeq 10^{9}\, \rm GeV$.

\vspace{0.4cm}
II) Dominance of non perturbative contributions from exotic instantons to perturbative terms: $M_{R}^{p}<<M_{R}^{E2}$. 

\vspace{0.3cm}
Under these assumptions, we counted the number of free-parameters as $13$. 
The (II) condition is exactly the definition of exotic see-saw mechanism: 
a see-saw mechanism completely dominated by exotic instantonic contributions. 

In this framework, acquiring 
11 low energy inputs from neutrino and quark physics, 
we demonstrated a successful leptogenesis 
for $M_{1}\simeq 3.5\times 10^{9}\, \rm GeV$,
$M_{2}\simeq M_{3}\simeq 8.7 \times 10^{9}\, \rm GeV$. 
and a $\theta_{13}\simeq 8.5^{0}$, 
while a precise dependence of the $m_{1}$ mass eigenvalues 
with the PMNS phase $\delta$ was shown in Fig.2 of Ref.\cite{Addazi:2015yna}.

\section{Gravitini as SHDM}

Assuming gravitini as Super Heavy LSP, 
we discuss bounds from CDM abundance. 
In particular, let us consider a scenario in which 
gravitini are so heavy to be produced not by thermal relic mechanisms
but by the non-thermal non-adiabatic expansion during inflation. 

This production mechanism is similar to every productions of pairs 
in an external background field. For example, an external electric 
field can promote a virtual electron-positron pair to a become 
real pair, as well as in Bekenstein-Hawking mechanism 
the external gravitational field can source B.H. pair 
to become a real pair nearby the black hole horizon. 
In our case, gravitini can be produced by the tremendous expansion rate 
during inflation of the FRW metric 
\begin{equation} \label{ds2}
ds^{2}=a^{2}(\eta)(d\eta^{2}-d{\bf x}^{2})
\end{equation}
For a massive heavy scalar field the action in FRW is easier than a spin $3/2$ field:
\begin{equation} \label{St}
S=\int dt \int d^{3}x \frac{a^{3}}{2}\left( \dot{\Phi}^{2}-\frac{(\nabla \Phi)^{2}}{a^{2}}-M_{X}^{2}\Phi^{2}-\zeta R \Phi^{2}\right)
\end{equation}
where $R$ is the Ricci scalar. 
Then, change the to time variable to the to conformal time variable
and we can perform a standard mode expansion of a QFT in curved space-time, 
\begin{equation} \label{Phi}
\Phi({\bf x})=\int \frac{d^{3}k}{(2\pi)^{3/2}a_{k}(\eta)}[a_{k}c_{k}(\eta)e^{i{\bf k}\cdot {\bf x}}
+a_{k}^{\dagger}c_{k}^{*}(\eta)e^{-i{\bf k}\cdot {\bf x}}]\end{equation}
where the coefficients of creation/destruction depend on the conformal time variable. 
This case was analyzed in Ref. \cite{Chung:1998zb}. 
However, our case is formally different and not still discussed in literature.

For a gravitino, the main dynamical aspects provoked by expansions are the same, 
but  the action has a Rarita-Schwinger fashion \cite{Schenkel:2011nv}:
\begin{equation}
\label{R}
S=\int d^{4}x e \bar{\psi}_{\mu}\mathcal{R}^{\mu}[\psi]
\end{equation}
where $\mathcal{R}$ is the Rarita-Schwinger operator:
\begin{equation}
\label{RS}
\mathcal{R}^{\mu}[\psi]=i \gamma^{\mu\nu\rho}\mathcal{D}_{\nu}\psi_{\rho}+m\gamma^{\mu\nu}\psi_{\nu}
\end{equation}
with the covariant derivative 
\begin{equation}
\label{cov}
\mathcal{D}_{\mu}\psi_{\nu}=\partial_{\mu}\psi_{\nu}+\frac{1}{4}\omega_{\mu ab}\gamma^{ab}\psi_{\nu}-\Gamma_{\mu\nu}^{\rho}\psi_{\rho}
\end{equation}
where $\gamma^{\mu_{1}...\mu_{n}}=\gamma^{[\mu_{1}}....\gamma^{\mu_{n}]}$
and $e={\bf det} e_{\mu}^{a}$ is the determinant of the inverse vielbein $e_{a}^{\mu}$. 

We can assume a torsion-free background metric so that $\Gamma_{\mu\nu}^{\rho}=\Gamma^{\rho}_{\nu\mu}$. 
Let us note that the mass is given by the K\"ahler and the superpotential 
as $m=e^{K/2}W/M_{Pl}^{2}$ and it can depedend by the space-time expansion. 
The related equation of motion is 
\begin{equation}
\label{eqRS}
(i  \mathcal{\not{D}}-m)\psi_{\mu}-(i\mathcal{D}_{\mu}+\frac{m}{2}\gamma_{\mu})\gamma \cdot \psi=0
\end{equation}
Now, in FRW, 
we have 
\begin{equation}
\label{cond}
e_{\mu}^{a}=a(\eta) \delta_{\mu}^{a},\,\,\,\,m=m(\eta),\,\,\,\, \omega_{\mu ab}=2\dot{a}a^{-1}e_{\mu[a}e_{b]}^{0}
\end{equation}

obtaining the EoM
\begin{equation}
\label{psiEOM}
i\gamma^{mn}\partial_{m}\psi_{n}=-(m+i\frac{a'}{a}\gamma^{0})\gamma^{m}\partial_{m}
\end{equation}
So that, we expand the gravitino field as 
\vspace{0.5cm}
\begin{equation}
\label{grav}
\psi_{\mu}(x)=\int \frac{d^{3}{\bf p}}{(2\pi)^{3}2p_{0}}\sum_{\lambda}\{ e^{i{\bf k}\cdot {\bf x}}c_{\mu}(\eta,\lambda)a_{k\lambda}(\eta)+e^{-i{\bf k}\cdot {\bf x}}c_{\mu}^{C}(\eta,\lambda)a_{k\lambda}^{\dagger}(\eta) \}
\ee
and note that the two coefficients are dependent in order to guarantee Majorana fermion condition $\psi^{C}=\psi$. 

At this point, we can perform a Bogoliubov transformation 
\begin{equation}
\label{cmuk}
c_{\mu k}^{\eta_{1}}=\alpha_{k}h_{k}^{\eta_{0}}(\eta)+\beta_{k}c_{k}^{C\eta^{0}}(\eta)
\end{equation}
so that we can calculate the energy density of gravitino produced:
\begin{equation}
\label{estimation}
\rho_{\psi}(\eta_{1})=mn_{\psi}(\eta_{1})=mH_{e}^{3}\left(\frac{1}{a(\eta_{1})} \right)\mathcal{I}
\end{equation}
$$\mathcal{I}=\int \frac{d k}{2\pi^{2}}k^{2}|\beta_{k}|^{2}$$
in which we consider Bogoliubov transformations 
from the Chauchy surface foliated by $\eta=\eta_{0}$
into another Chauchy surface in a cosmological time 
$\eta_{1}>\eta_{0}$, assuming conditions 
$\dot{a}/a^{2}<<1$, and we tacitly normalized 
$k\rightarrow k/a H_{e}$, $\eta\rightarrow \eta a_{e}H_{e}$, 
$a\rightarrow a/a_{e}$, where $e$-label 
indicates variables of the starting oscillating inflaton epoch.

After that, it is crucially important to constrain the ragion
of parameters from CDM abundance. 
The relation among gravitini energy density and radiation is 
\begin{equation}
\label{gravitino}
\frac{\rho_{\tilde{G}}(t_{0})}{\rho_{R}(t_{0})}=\frac{\rho_{\tilde{G}}(t_{Re})}{\rho_{R}(t_{Re})}\left(\frac{T_{R}}{T_{0}} \right)
\end{equation}
where $t_{0}$ is the today time, while the ratio
$\rho_{\tilde{G}}(t_{Re})/\rho_{R}(t_{Re})$
is determined after the Reheating epoch, 
while gravitini were produced during $t_{e}>t_{Rh}$
epoch, where inflaton starts to oscillate 
and decays into SM particles. 
During the radiation dominated epoch, the ratio $\rho_{\tilde{G}}(t_{Re})/\rho_{R}(t_{Re})$
can be estimated as 
\begin{equation}
\label{XR}
\frac{\rho_{\tilde{G}}(t_{Rh})}{\rho_{R}(t_{Rh})}\simeq \frac{8\pi}{3}\left(\frac{\rho_{\tilde{G}}(t_{0})}{M_{Pl}^{2}H^{2}(t_{0})} \right)
\end{equation}
where $H_{0}\simeq 100\, \rm h\,km\, s^{-1}\, Mpc^{-1}$. 
Then, the natural hierarchies of Hubble scales and densities with inflaton parameters are
$H^{2}(t_{e})\sim m_{\phi}^{2}$ and $\rho(t_{e})\sim m_{\phi}^{2}M_{Pl}^{2}$, 
so that 
\begin{equation}
\label{begine}
\Omega_{\tilde{G}}h^{2}\sim 10^{17}\, \left( \frac{T_{Rh}}{10^{9}\, \rm GeV}\right)\left(\frac{\rho_{\tilde{G}}(t_{e})}{\rho_{c}(t_{e})} \right)
\end{equation} 
where $\rho_{c}=3H(t_{e})^{2}M_{Pl}^{2}/8\pi$ is the critical energy density during $t_{e}$. 
Eq.(\ref{begine}) can also be rewritten as 
\begin{equation}
\label{rewe}
\Omega_{\tilde{G}}h^{2}\simeq \Omega_{R}h^{2}\left(\frac{T_{Rh}}{T_{0}} \right)\frac{8\pi}{3}\left( \frac{M_{\tilde{G}}}{M_{Pl}}\right)\frac{n_{\tilde{G}}}{M_{Pl}H^{2}(t_{e})}
\end{equation}
But the limit on the inflaton mass is $m_{\phi} \simeq 10^{13}\, \rm GeV$ or so, while 
the limit on $T_{Rh}$ for a successful  reheating is 
$T_{Rh}/T_{0}\simeq 4.2\times 10^{14}$, for $\Omega_{\tilde{G}}h^{2}\lesssim 1$
In this range of parameters, 
we are able to constrain the Gravitino mass. 
How heavy might it be?
We can set the limit 
$$\,\,\,\,\,\,\,\,\,m_{\tilde{G}}\simeq (10^{-2}\div 2)\times H\simeq 10^{11}\div 10^{13}\, \rm GeV.$$
This sets a limit on the SUSY symmetry breaking scale 
of 
$$M_{SUSY}>m_{\tilde{G}}\simeq 10^{11} \div10^{13}\, \rm GeV,$$ 
that is well compatible with our D-brane leptogenesis bound $M_{SUSY}>10^{9}\, \rm GeV$. 
More precisely, not only the supersymmetric scale has to satisfy this hiearachy with the gravitino mass, 
but we cannot allow any superparticles to be smaller than the gravitino, 
otherwise gravitini will suddenly decay on them. 
Let us not that this also set a bound on the sting scale and on the size of the 3-cycles
wrapped by the $E2$-brane responsible for RH neutrino masses. 
In particular, stability of exotic instanton calculations 
are trustable if $M_{S}\geq M_{SUSY}\simeq 10^{11}\div 10^{13}\, \rm GeV$.
The order of neutrino masses are $M_{1,2,3,RH}\simeq e^{-S_{E2}(\rho_{i})}M_{S}\simeq 10^{9}\, \rm GeV$, 
where $e^{-S_{E2}(\rho_{i})}$ is the non-perturbative coupling constant associated to the 
exotic instanton. In particular, it depends on geometric moduli fields $\rho_{i}$, parametrizing the size of
the 3-cycles wrapped by the $E2$-brane in the internal Calabi-Yau compactification $CY_{3}$. 
So that, this imposes an interesting bound on the $E2$-brane: 
$e^{-S_{E2}}\simeq M_{RH,1,2,3}/M_{S}<10^{-2}\div 10^{-3}$.

\section{Conclusions and discussions}

In this paper, we discussed the gravitino problem in contest of
a see-saw model for neutrini and leptogenesis 
strongly motivated by IIA open superstring theory.
In particular, 
we suggest an effective Pati-Salam (un-)oriented quiver field theory,
 UV completed by
an intersecting D-brane superstring theory. 
The quiver theory encodes informations on D-branes and its consistency 
with respect to gauge and stringy anomalies or tadpoles. 
This model assumes supersymmetry to be spontaneously broken 
at very high scale: $M_{SUSY}>10^{9}\, \rm GeV$. 
We individuated a possible problem on this assumption: 
if the gravitino is heavy and unstable, it will decay inside the thermal bath and our leptogenesis scenario 
will not be trustable! Essentially gravitini decays will jeopardize the leptogenesis 
picture, badly washing-out the lepton-antilepton asymmetry generated by RH neutrini decays. 
So that, we moved ourself to try a way-out on this problem in contest of 
our model. 
We not only demonstrated that a way-out is possible, but 
we also provide a related candidate for cold dark matter. 
In particular, we noticed that if the gravitino is the Lightest Stable Particle 
of the Superworlds, then the gravitino bound is automatically avoided:
gravitini simply cannot decay into more massive superparticles. 
On the other hand, gravitini are expected to be very massive 
and they can provide a good candidate for Super Heavy Dark Matter
or WIMPZILLA. In particular, we re-discussed the 
gravitational production of massive gravitini during the inflation 
 due to the 
tremendous expansion rate. 
We show that gravitino mass has to be around the inflaton mass or so 
in order to guarantee a correct abundance of CDM:
$10^{11}\div 10^{13}\, \rm GeV$. 
This also sets a limit on the instantonic coupling of $e^{-S_{E2}}<10^{-2}\div 10^{-3}$. 
So that, we tried a relation of the mechanism of DM production with 
parameters of neutrino and quark physics. 
So, we conclude that our model naturally provides a coherent picture of baryogenesis
and dark matter genesis: the parameters of the two genesis mechanisms are rigidly related each others. 
Our model inspired by first principles of superstring theory naturally 
reproduces and relates ordinary and dark matter. 
On the other hand, we can relax the hypothesis that all cold dark matter is composed of superheavy gravitini
\footnote{This assumption can be strongly motivated by unifying picture of dark matter and dark energy by 
hidden gauge sectors recently suggested in Refs. \cite{Addazi:2016sot,Addazi:2016nok}
as well as by DAMA dark matter direct detection result
recently analyzed in a fully detailed analysis in Ref.\cite{Addazi:2015cua},
in the framework of asymmetric Mirror dark matter.
 }.
In this case, the bound on the gravitino mass set by the non-adiabatic genesis mechanism 
can be relaxed as well as constrains on the instantonic coupling. 
Let us also comment about possible implications in phenomenology.
Clearly, such a candidate is so heavy and rarely distributed in the galactic 
halo that cannot be seen by direct detection experiments. 
Such a candidate could be seen by high energy DM indirect detection.
It is also possible that gravitini are destabilized by trilinear R-parity breaking terms like $y_{hL}H_{\alpha} L^{\alpha}$
so that it can decay as $\tilde{G}\rightarrow \gamma \nu$, with 
a rate \cite{Takayama:2000uz}
$$\Gamma(\tilde{G}\rightarrow \gamma \nu)=\frac{\cos^{2}\theta_{W}}{32\pi}\frac{m_{\nu}}{m_{\chi}}\frac{m_{\tilde{G}}^{3}}{M_{Pl}^{2}}\left(1-\frac{m_{\nu}^{2}}{m_{\tilde{G}}^{2}} \right)^{3}\left(1+\frac{m_{\nu}^{2}}{3 m_{\tilde{G}}^{2}} \right)$$
induced by an interaction term 
$$L_{int}=-\frac{i}{8M_{Pl}}\bar{\psi}_{\mu}[\gamma^{\nu},\gamma^{\rho}]\gamma^{\mu}\lambda F_{\nu\rho}$$
where $\lambda$ is the photino and $\chi$ is the bino-dominated lightest neutralino eigenstate. 
In other words, the photino and the higgsino are rotated into neutralino so that 
and effective neutralino-neutrino mixing mediates the gravitino decays into a neutrino and a photon. 
But, 
assuming $m_{\chi}\simeq 10^{13}\, \rm GeV$
and $m_{\tilde{G}}\simeq 10^{11}\, \rm GeV$, 
we obtain 
$\Gamma\simeq 10^{-24} \times 10^{-16} \times 10^{11}\, \rm GeV \simeq 10^{-20}\, \rm eV$
corresponding to $\tau \simeq 10^{5}\, \rm s$ that is clearly a too fast rate for 
a good dark matter candidate. 
At least, considering a very heavy $\chi$-particle, lets say $10^{15}\, \rm GeV$,
we can enhance $\tau \simeq 1\,\rm yr$ or so. 
But these R-parity violating perturbative terms are expected to be suppressed by non-perturbative stringy effects,
as discussed in section II. 
So that, the gravitino decay can be only induced by higher order effective operator
like 
$\frac{S^{n}}{\Lambda^{n}}h_{\alpha}L^{\alpha}$
suppressing the rate of at least a factor $10^{-13}$ for $m_{\chi}\simeq 10^{13}\, \rm GeV$ in order to obtain 
a cosmological lifetime $1\div 10\, \rm Gyrs$, 
where $\Lambda$ can be for example the non-perturbative scale of a R-R or NS-NS flux in the compactification. 
For example for supposing $n=1$, the rate will be suppressed as $(\langle S \rangle/\Lambda)^{2}\simeq 10^{-13}$.
Several different operators involving other heavier Higgs in Pati-Salam could be generated by non-perturbative stringy effects
as understood. These terms can be generated in our model by expectation values of non-perturbative R-R or NS-NS fluxes. 
To find a $10^{11}\div 10^{13}\, \rm GeV$ neutrino or photon 
with two-body decay peaks would provide a test-bed in favor of our 
suggestion. Experiments like IceCube and ANTARES 
would search for such a very rare decays.
Of course, other possible mechanism of DM genesis can be considered. 
For example, in 
  \cite{Khlopov:1985jw,Khlopov:2004tn,Khlopov:2008qy}, several different DM genesis mechanisms from Bekenstein-Hawking evaporation of primordial black holes 
were suggested. A complete study of these mechanisms in contest of our model 
deserves further investigations beyond the purposes of this letter
\footnote{We mention that recently we suggested a possible way-out to the information paradox related to Bekenstein-Hawking radiation,
relating it to quantum chaotization of infalling information \cite{Addazi:2015cho,Addazi:2015hpa,Addazi:2016cad}.
On the other hand, we found in contest of $f(R)$-gravity that primordial
Nariai black holes have an antievaporation instability, turning-off Bekenstein-Hawking radiation. 
So that in these models, a DM production from an evaporation of primordial BH seems to be not 
possible \cite{Addazi:2016prb}.}.

\vspace{2cm} 

{\large \bf Acknowledgments} 
\vspace{3mm}

AA would like to thank Massimo Bianchi and Giulia Ricciardi for valuable discussions on these subjects.
AA would also like to thank LMU and Fudan University, for hospitality during the preparation of this letter. 
AA work was supported in part by the MIUR research grant Theoretical Astroparticle Physics PRIN 2012CP-PYP7 and by SdC Progetto speciale Multiasse La Societ\'a della Conoscenza in Abruzzo PO FSE Abruzzo 2007-2013.
The work by MK was performed within the framework of the
Center FRPP supported by MEPhI Academic Excellence
Project (contract 02.03.21.0005, 27.08.2013), in which
the part on initial cosmological conditions was supported
by the Ministry of Education and Science of Russian Federation,
project 3.472.2014/K and on the forms of dark
matter by grant RFBR 14-22-03048.

\end{document}